\documentstyle[prl,aps,epsfig,twocolumn]{revtex}

\begin{document}
\draft

\twocolumn[\hsize\textwidth\columnwidth\hsize\csname@twocolumnfalse\endcsname
\title{Adiabatic passage of collective excitations in atomic ensembles}
\author{Y. Li, Y. X. Miao, and C. P. Sun}
\address{Institute of Theoretical Physics, The Chinese
Academy of Sciences, Beijing 100080, China}
\maketitle
\begin{abstract}
  We describe a theoretical scheme that allows for transfer of quantum
states of atomic collective excitation between two macroscopic
atomic ensembles localized in two spatially-separated domains. The
conception is based on the occurrence of double-exciton dark
states due to the collective destructive quantum interference of
the emissions from the two atomic ensembles. With an adiabatically
coherence manipulation for the atom-field couplings by stimulated
Ramann scattering, the dark states will extrapolate from an
exciton state of an ensemble to that of another. This realizes the
transport of quantum information among atomic ensembles.
\end{abstract}
\pacs{PACS number: 03.67.-a, 71.36.+c, 03.65.Fd, 05.30.Ch,
42.50.Fx} \vspace {1cm}]

\section{Introduction}

In quantum information processing (QIP), the physical implementations of
quantum memories and quantum carriers can be considered as one of the
biggest challenges for physicists. Since photons are manifested to be ideal
carriers of quantum information in many experiments, more attentions are
paid to investigations on ideal storage systems of quantum information
recently. In usual, quantum memories not only allow the transport of unknown
quantum states among separated locations with short access times \cite%
{PhysQI}, but also possess the nature avoiding quantum decoherence, namely,
one can isolate them completely from environmental interactions and control
their coupling to a necessary extent. To satisfy these basic requirements,
the collective excitations of symmetric internal states of an atomic
ensemble \cite{RMP} have been studied extensively as a very attractive
candidate for this in last years\cite{lukin1,Flei0,Flei1,zol0}. For this
quantum memory realized by atomic ensemble, despite the seemingly
insurmountable difficulties (such as the quantum leakage due to the
inhomogeneous coupling of external fields \cite{sun-yi-you} and quantum
decoherence induced by the center of mass (C.M.) motion\cite{li-yi-you-sun}%
), there may be a hope to overcome them by considering that the existence of
equivalence classes of collective storage states will reduce the enhanced
sensitivity of the collective memory to environmental interactions.

In this paper we will work on the problem how to transport quantum
information between two atomic ensembles in direct way other than
the current aspect how to store quantum information. As an
indirect way to solve this problem, a significant contribution by
Duan {\it et al.} \cite{zoll} is that, using two un-correlated
linear polarized lights to couple two non-correlated atomic
ensembles, the entanglement between the two atomic ensembles can
be created after a non-local Bell measurement for the circular
polarization mode based on the Stokes variables mixing linear
polarization mode. Then, the quantum communication between two
atomic ensembles can be implemented. Recent experimental success
\cite{polzik} clearly demonstrates the power of such an atomic
ensemble based system for entangling macroscopic objects as the
so-called canonical teleportation defined for arbitrary pair of
canonical variables\cite{sun-yu} beyond the usual coordinate and
momentum. Our direct way to communicate quantum information
between two atomic ensembles does not depend on any post-selective
measurement.

The quantum memories in our protocol are also the certain collective quantum
states of two atomic ensembles, which describe the collective low
excitations of two clusters of atoms confined in two well-separated well
potentials. It is found that, for the two excited atom systems interacting
with a single mode quantized light field, there exist double-exciton dark
states decoupling with light field, which extrapolate from an exciton state
of an ensemble to that of another. To change the effective couplings of
atomic ensembles to the light field adiabatically through a Ramann light
stimulation\cite{ramann}, one can transfer a single exciton state of one
ensemble to another. In mathematical formulation, the double-exciton dark
state is quite similar to the polariton state in the atomic ensemble storage
scheme of photon Fock state \cite{lukin1,Flei0,Flei1,zol0}, but the physical
difference is rest on that our double-exciton dark state does not contain
any variable of light photon.

This paper is organized as follows. In section 2, we propose our protocol of
transfer of the collective excitations of atomic ensembles and deduce its
corresponding model Hamiltonian in a simplified way. The transfer of quantum
information in terms of collective atomic states is depicted in section 3 as
the time evolution governed by the adiabatically manipulated Hamiltonian. In
section 4 the exact solution to the adiabatic dynamics is given in the
macroscopic limit by defining the double-exciton dark states.

\section{Model for Two Atomic Ensembles with Two Tunable Couplings}

In our protocol the total system (Fig. 1) involving two atomic ensembles is
localized in two far-separated places, the left and right one containing $%
N_l $ and $N_r$ two-level atoms respectively. These two clusters of atoms
are coupled to a light field with coupling coefficients $g_l$ and $g_r$. To
implement our scheme, it is most important for the experimental setup to
require that the effective strengths $g_l$ and $g_r$ of light field coupling
can be varied relatively or independently. This requirement shows the
drawback of a naive scheme using the practical two-level atom with direct
coupling between atom and light field: since the same single mode light
interacts with two atomic ensembles at the same time, one can not change the
relative effective coupling strengths of field with two atomic ensembles.
Thus, it is impossible to reach our goals by changing the coupling directly.
In the following we will use the equivalent two-level system deduced from
the practical three-level atoms stimulated by Ramann scattering.

%
\begin{figure}[h]
\includegraphics[width=5.5cm,height=3cm]{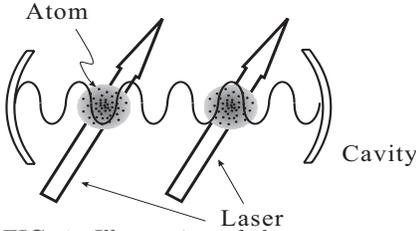}
\caption{Illustration of the system: two 3-level atomic ensembles consisted
of the identical atoms are located in a quantized cavity and coupled to two
classical laser field, respectively.}
\end{figure}

For conceptual simplicity we here assume that each atom used in practice
only has three energy levels as shown in Fig.~2. There are two meta-stable
lower states ($|g\rangle $ and $|e\rangle $) and an auxiliary state $%
|a\rangle $. The transition $|a\rangle \rightarrow |g\rangle $ of each of
these atoms in both wells is coupled to a same quantized radiation mode with
Rabi-frequency $\Omega $ and frequency $\omega $ while the transitions $%
|a\rangle \rightarrow |e\rangle $ of the atoms in the left (right) well are
driven by a classical control field of Rabi-frequency $\Omega _l$ ($\Omega
_r $). Moreover, we also assume that the detuning between $|g\rangle $ and $%
|a\rangle $ with respect to the quantized light is the same as that between $%
|e\rangle $ and $|a\rangle $ with respect to the classical light. Then, we
can write down the single atom Hamiltonians in the interaction picture
\begin{equation}
H_{Is}=[\Omega |g(s)\rangle a^{\dagger }+\Omega _s|e(s)\rangle ]\langle
a(s)|e^{-i\Delta t}+h.c.
\end{equation}
for $s=l,r$. Here, $a^{\dagger }$ is the creation operator of the quantized
field. Due to the stimulated Ramann effect for large detuning $\Delta $, the
effective Hamiltonians can be obtained as
\begin{eqnarray}
H_s &=&-\frac{|\Omega |^2}\Delta a^{\dagger }a|g(s)\rangle \langle g(s)|-%
\frac{|\Omega _s|^2}\Delta |e(s)\rangle \langle e(s)|  \nonumber \\
&&-[\frac{\Omega \Omega _s^{*}}\Delta |g(s)\rangle \langle e(s)|a^{\dagger
}+h.c.]
\end{eqnarray}
by an adiabatic elimination of the upper level $|a\rangle $. This just
realizes an equivalent two-level atom system with the excited state $%
|e\rangle $ and ground state $|g\rangle $ with effective coupling
\begin{equation}
g_s=-\frac{\Omega \Omega _s^{*}}\Delta ,\text{ }s=r,l.
\end{equation}
$g_s$ can be set real by adding a proper phase to $|g(s)\rangle $ or $%
\langle e(s)|.$ In this sense, the effective coupling strengths $g_l$ and $%
g_r$ or their ratio (relative effective coupling strength) can be well
controlled independently. If the level difference between $|e\rangle $ and $%
|g\rangle $ is $\omega _a$, the effective level differences of the
equivalent two-level atoms in the left and right well are
\[
\omega _s=\omega _s(a^{\dagger }a)=\omega _a+\frac{|\Omega |^2}\Delta I-%
\frac{|\Omega _s|^2}\Delta ,
\]
where $s=r,l$ and $I=a^{\dagger }a$ is regarded as the density of the
quantized light. In most cases we can neglect the Stark shifts $\frac{%
|\Omega |^2}\Delta $ and $\frac{|\Omega _s|^2}\Delta $.

Generalization to multi-mode field and multi-level atoms is straightforward.
Let the left and right well contain $N_l$ and $N_r$ atoms respectively.
Then, the many-atom Hamiltonian is given by

\begin{eqnarray}
H &=&\frac 12\omega _l\sum_{j=1}^{N_l}\sigma _z^{[j]}(l)+\frac 12\omega
_r\sum_{j=1}^{N_r}\sigma _z^{[j]}(r)+  \nonumber \\
&&\omega _aa^{\dagger }a+\{a[\sum_{j=1}^{N_l}g_l\sigma
_{+}^{[j]}(l)+\sum_{j=1}^{N_r}g_r\sigma _{+}^{[j]}(r)]+h.c.\}.
\end{eqnarray}
Here, we have defined the quasi-spin ladder operators $\sigma
_{+}^{[j]}(s)=|e(s)\rangle _{jj}\langle g(s)|$ and $\sigma
_{-}^{[j]}(s)=[\sigma _{+}^{[j]}(s)]^{\dagger }$ and the population
inversion operator $\sigma _z^{[j]}(s)=|e(s)\rangle _{jj}\langle
e(s)|-|g(s)\rangle _{jj}\langle g(s)|$ in terms of the excited and ground
states $|e(s)\rangle _j$ and $|g(s)\rangle _j$ $(s=l,r)$ in the two wells.
It is noticed that, for simplicity, the coupling constants $\Omega ,\Omega
_l $ and $\Omega _r$ between the atoms and the fields are assumed to be
equal for all atoms in the well.

%
\begin{figure}[h]
\includegraphics[width=7cm,height=4cm]{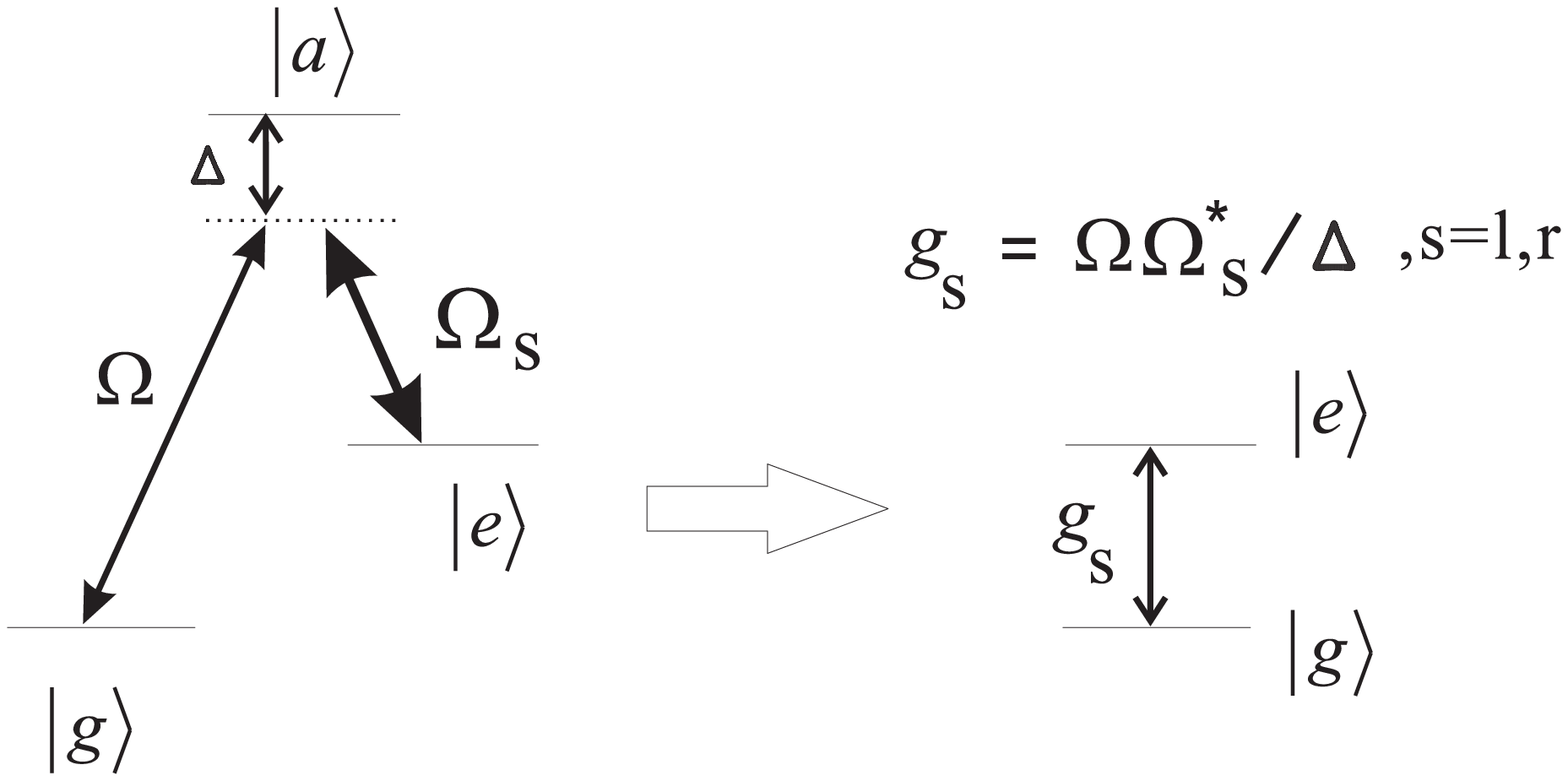}
\caption{Illustration of the scheme for adiabatic transport of collective
excitations. The right figure shows the equivalent two-level atom induced by
Ramann Scattering of a three-level atom.}
\end{figure}

When all atoms are prepared initially in the symmetric ground state $|{\bf g}%
_{l}\rangle $ $=|g_{1}(l),g_{2}(l),\dots g_{N_{l}}(l)\rangle $ and $|{\bf g}%
_{r}\rangle $ $=|g_{1}(r),g_{2}(r),\dots g_{Nr}(r)\rangle ,$ the $n$th
excited state of the system can be described by symmetrizing the effects of $%
n$ flips of different atoms from $|g_{j}(s)\rangle \rightarrow
|e_{j}(s)\rangle $. In fact, they are the totally symmetric Dicke-states
\[
|{\bf e(}s{\bf )}\rangle =\frac{1}{\sqrt{N_{s}}}\sum_{j=1}^{N_{s}}|g_{1}(s),%
\dots ,e_{j}(s),\dots \rangle ,
\]%
\begin{eqnarray}
|{\bf e}^{2}{\bf (}s{\bf )}\rangle
&=&\frac{1}{\sqrt{2N_{s}(N_{s}-1)}}
\nonumber \\
&&\sum_{i\neq j=1}^{N_{s}}|g_{1}(s),\dots
e_{i}(s),...,e_{j}(s),\dots \rangle ,
\end{eqnarray}

\[
.....................
\]

\bigskip

\section{Adiabatic Transfer of Collective Excitations}

To get the basic idea for our quantum state transferring mechanism via
adiabatic manipulation, we first consider the special case of the above
model: there is only one atom in each well. For such a system of two atoms
interacting with a single mode light field, there exists an asymmetric
two-atom state, the dark state
\begin{equation}
|d(g_{l},g_{r})\rangle \sim \cos \theta |e(l),g(r)\rangle -\sin \theta
|g(l),e(r)\rangle
\end{equation}%
decoupling from the light {\rm vacuum} field, which is a special eigenstate
of the atom-field system \cite{dicke}. Here, the mixing angle is defined by $%
\tan \theta =\frac{g_{l}}{g_{r}}$. Due to the destructive quantum
interference between the two transitions from the two atoms in this state,
the so-called emission trapping occurs, namely, no emission transmits from
such a state. Furthermore, by changing the coupling strengths $g_{l}$ and $%
g_{r}$ sufficiently slowly, the mixing angle $\theta $ can be rotated from $%
0 $ to $\pi /2$. According to the quantum adiabatic theorem, if the total
system is initially prepared in the dark state $|d(g_{l},g_{r})\rangle $,
the adiabatic following will determine the quantum state transfer from $%
|e(l),g(r)\rangle $ to $-|g(l),e(r)\rangle $. This mechanism for adiabatic
transfer is the basis of our scheme to transport quantum states from an
atomic ensemble to another.

Motivated by the above conceptions for the two-atom case, the immediately
following problem is to find the dark state for the many-atom case, we write
down the collective quasi-spin operators:
\begin{equation}
S_\beta (s)=\sum_{j=1}^{N_s}\sigma _\beta ^{[j]}(s)
\end{equation}
($\beta =\pm ,z;s=l,r$). Correspondingly the mixing collective operators

\begin{equation}
\varphi =\,\frac{\cos \alpha }{\sqrt{N_l}}S_{-}(l)-\,\frac{\sin \alpha }{%
\sqrt{N_r}}S_{-}(r)
\end{equation}
and

\begin{equation}
\phi =\frac{\sin \alpha }{\sqrt{N_{l}}}S_{-}(l)+\,\frac{\cos \alpha }{\sqrt{%
N_{r}}}S_{-}(r)
\end{equation}%
can be defined to depict the double excitation in two atomic ensembles where
the mixing angle $\alpha $ is defined by $\tan \alpha \,=\frac{g_{l}\sqrt{%
N_{l}}}{g_{r}\sqrt{N_{r}}}$. In terms of these mixing collective operators,
we can define the double excitation of the two atomic ensembles by
\begin{equation}
|D(n)\rangle =|D(n,g_{l},g_{r})\rangle =\frac{1}{\sqrt{n!}}\varphi ^{\dagger
n}{\bf |g_{l},g_{r}\rangle ,}
\end{equation}%
where the ground state ${\bf |g_{l},g_{r}\rangle =}|{\bf g}_{l}\rangle $ $%
\otimes |{\bf g}_{r}\rangle \otimes |0\rangle _{p}$ is the total ground
state, ($|0\rangle _{p}$ shows that there is $0$ photon in the light {\rm %
vacuum} field) and
\begin{equation}
|{\bf e_{l}^{m},e_{r}^{n}\rangle }{\bf =}\frac{1}{\sqrt{%
N_{l}^{m}N_{r}^{n}m!n!}}S_{+}^{m}(l)S_{+}^{n}(r){\bf |g_{l},g_{r}\rangle }
\end{equation}%
represents the $m$ atom excitations in the left ensemble and $n$ atom
excitations in the right one (with $0$ photon in the light {\rm vacuum}
field) with $|{\bf e_{l}^{m},e_{r}^{n}\rangle =}|{\bf e}^{m}{\bf (}l{\bf %
)\rangle \otimes }|{\bf e}^{n}(r){\bf \rangle \otimes }|0\rangle _{p}$. Now
we can show the central result\ that $|D(n)\rangle $ is approximately a dark
many-atom state, namely, it is cancelled by the interaction Hamiltonian
\[
H_{I}=g(a\phi ^{\dagger }+a^{\dagger }\phi )
\]%
or
\[
H_{I}|D(n)\rangle \simeq 0,
\]%
where $g=\sqrt{g_{l}^{2}N_{l}+g_{r}^{2}N_{r}}$. In fact, it is easy to check
the commutation relation
\[
\lbrack \varphi ^{\dagger },\phi ]=\frac{1}{2}\sin 2\alpha \lbrack \frac{%
S_{z}(l)}{N_{l}}-\frac{S_{z}(r)}{N_{r}}]
\]%
and the limit behaviors
\[
\frac{S_{z}(l)}{N_{l}}\rightarrow -1,\frac{S_{z}(r)}{N_{r}}\rightarrow -1
\]%
for very large $N_{l}$ and $N_{r}$ in the low excitation case that most
atoms occupy the ground state. Then, $[\varphi ^{\dagger },\phi ]\rightarrow
0$ and thus
\[
H_{I}|D(n)\rangle \sim \phi \varphi ^{\dagger n}{\bf |g_{l},g_{r}\rangle }%
\sim \varphi ^{\dagger n}\phi {\bf |g_{l},g_{r}\rangle }=0{\bf .}
\]%
\

With the above observations, we can describe our scheme about the transport
of quantum information carried by the quantum state of atomic ensemble in a
quite similar way to the well-known scheme for the storage of photon
information by atomic ensemble. For the resonance case in the interaction
picture, as the eigenstates of $H_{I}$ with zero eigenvalue, the dark states
$|D(n)\rangle $ are degenerate for $n=0,1,2,...$. Adiabatic changes of the
two classical light fields through $\Omega _{l,r}$ can induce the
independent variances of the effective coupling constants $g_{l}$ and $g_{r}$%
. This can result in the asymptotic behavior of the dark states by varying
the mixing angle $\alpha $ from $0$ to $\pi /2$, which physically
corresponds to two limit cases:
\[
g_{r}\sqrt{N_{r}}\gg g_{l}\sqrt{N_{l}};g_{l}\sqrt{N_{l}}\gg g_{r}\sqrt{N_{r}}%
.
\]%
In this process, the number of excitons in the dark state $|D(n)\rangle $ is
an adiabatic invariant and thus the adiabatic following along $|D(n)\rangle $
leads to the collective quantum state varies from
\[
|D(n,g_{l}=0,g_{r}=1)\rangle =|{\bf e_{l}^{n},e_{r}^{0}\rangle =}|{\bf e}^{m}%
{\bf (}l{\bf )\rangle }\otimes |0\rangle _{p}
\]%
to
\[
|D(n,g_{l}=1,g_{r}=0)\rangle =|{\bf e_{l}^{0},e_{r}^{n}\rangle =}|{\bf e}^{n}%
{\bf (}r{\bf )\rangle }\otimes |0\rangle _{p}.
\]%
If the initial quantum information stored in the left ensemble is described
by the density matrix $\hat{\rho}_{l}=\sum_{n,m}\rho _{nm}\,|{\bf e}^{n}{\bf %
(}l{\bf )\rangle }\langle {\bf e}^{m}{\bf (}l{\bf )}|$, the adiabatic
transfer process generates a same quantum state of collective excitations in
the right atomic ensemble $\hat{\rho}_{r}=\sum_{n,m}\rho _{nm}\,|{\bf e}^{n}%
{\bf (}r{\bf )\rangle }\langle {\bf e}^{m}{\bf (}r{\bf )}|$. In fact, the
adiabatic evolution of the total state
\begin{equation}
\hat{\rho}(g_{l},g_{r})=\sum_{n,m}\rho _{nm}|D(n,g_{l},g_{r})\rangle \langle
D(m,g_{l},g_{r})|
\end{equation}%
in the interaction picture just extrapolates between two factorized states
\begin{equation}
\hat{\rho}(g_{l}=0,g_{r}=1)=\hat{\rho}_{l}\otimes |{\bf g}_{r}\rangle
\langle {\bf g}_{r}|\otimes |0\rangle _{pp}\langle 0|
\end{equation}%
and
\begin{equation}
\hat{\rho}(g_{l}=1,g_{r}=0)=|{\bf g}_{l}\rangle \langle {\bf g}_{l}|\otimes
\hat{\rho}_{r}\otimes |0\rangle _{pp}\langle 0|.
\end{equation}%
Since $\hat{\rho}_{l}$ and $\hat{\rho}_{r}$ possess the same single exciton
density matrix, one may say the quantum information has been transported
from the left atomic ensemble to the right one.

Like the problem in the scheme of atomic ensemble storage of photon quantum
information \cite{hau}, there still exist two problems, quantum decoherence
due to the interaction with environment and quantum leakage due to
inhomogeneous coupling of the classical (and quantum) fields \cite%
{sun-yi-you},\cite{li-yi-you-sun}. The former can be partially
solved by considering that there are equivalence classes of
collective storage states besides the non-degenerate symmetric
state with maximum $J_s=\frac{N_s}2$ and the transitions to them
do not affect the reading-out of collective quantum state of
exciton in right ensemble to which the quantum information is
transferred. The fact was found recently by Fleischhauer {\it et
al} as a key element to overcome decoherence in atomic ensemble
quantum information processing \cite{Flei0}. Our present model can
be used to illustrate this interesting idea by the introduction of
non-symmetric single excitation states
\begin{equation}
|{\bf e_k(}s{\bf )}\rangle =\frac 1{\sqrt{N_s}}\sum_{j=1}^{N_s}e^{\frac{%
2i\pi jk}{N_s}}|g_1(s),\dots ,e_j(s),\dots ,g_{N_s}(s)\rangle
\end{equation}
for $k=0,1,2,..,N_s-1$ and the corresponding collective excitation operators
\begin{equation}
S_{\beta k}(s)=\frac 1{\sqrt{N_s}}\sum_{j=0}^{N_s-1}e^{\frac{2i\pi jk}{N_s}%
}\sigma _\beta ^{[j]}(s)
\end{equation}
for $k=0,1,2,...N_s-1$ $(\beta =\pm ,z;s=l,r).$ For very large $N_s$, we
have a complete set
\begin{equation}
\{b_k(s)=\lim_{N_s\rightarrow \infty }\frac 1{\sqrt{N_s}}%
S_{-k}(s)|k=0,1,..N_s-1\}
\end{equation}
of boson operators. There are $N_s$ fault-tolerant equivalence classes $C_n:$%
\begin{equation}
\{\prod_{k=0}^{N_s-1}b_k^{\dagger m_k}(s)\varphi ^{\dagger n}|0\rangle
|m_k=0,1,...N_s-1\}
\end{equation}
which contain $N_s^2-1$ elements besides the non-degenerate
symmetric state with maximum $J_s$. The transporting process of
the quantum state does not distinguish among those states. A
physical explanation of these equivalence classes is given in
terms of quasi-particle excitations. Only excitations of double
dark-state exciton modes with specific wave vectors couple to the
quantum field. Excitations of other excitons by $b_k^{\dagger
m_k}$ do not affect the quantum state transporting. Hence, from
the point of view of quantum memory, all collective atomic states
with the same number of $\varphi ^{\dagger }$ excitations are
equivalent. Then, taking this equivalence into account, one can
see that the equivalence class will
compensate the $\sqrt{N_s}$ enhancement of decoherence \cite{sun-yi-you},%
\cite{li-yi-you-sun}.

\section{\protect\bigskip Quantum Dynamics in Macroscopic Limit}

In the macroscopic limit with very large atomic numbers $N_{l}$ and $N_{r}$,
the two central results in this paper will be given in this section: (1) $%
\varphi $ and $\phi $ define bosonic excitations cooperating between the two
atomic ensembles. (2) The collective Fock state $|D(n)\rangle $ becomes a
dark state decoupling from the quantized light field.

To illustrate them, we first consider the general observation about the
classical limit of the quantum angular momentum operators ${\bf J}%
=(J_1,J_2,J_3)$. It was already noticed that, when the total angular
momentum $J$ approaches infinite, algebra $su(2)$ gives a representation of
the usual bosonic Heisenberg algebra by defining the boson operators
according to $b=\lim_{J\rightarrow \infty }\frac{J_1-iJ_2}{\sqrt{2J}}$,\quad
$b^{\dagger }=\lim_{J\rightarrow \infty }\frac{J_1+iJ_2}{\sqrt{2J}}$. The
above limit is taken under the low excitation condition that $J-J_3$ is
finite. The detailed argument was given recently for the exciton-polariton
problem in semiconductor microcavity \cite{sun-liu}. Considering that $%
S_\beta (s)=\sum_{j=1}^{N_s}\sigma _\beta ^{[j]}(s)$ gives a spinor
representation of angular momentum with $J_s=\frac{N_s}2,$ the {\em %
classical limit} of angular momentum can be realized as the {\it macroscopic
limit }with $N_s\rightarrow \infty .$ In this limit, we have the independent
excitations described by bosonic operators

\begin{equation}
b_l=\lim_{N_l\rightarrow \infty }\frac{S_{-}(l)}{\sqrt{N_l}}%
,b_r=\lim_{N_r\rightarrow \infty }\frac{S_{-}(r)}{\sqrt{N_r}}.
\end{equation}
Then the above defined double collective excitation can be described by the
canonical transformations
\begin{eqnarray}
\varphi &=&b_l\cos \alpha \,-b_r\sin \alpha ,  \nonumber \\
\phi &=&b_l\sin \alpha \,\,+b_r\,\cos \alpha ,  \label{phi,psi}
\end{eqnarray}
of two boson modes $b_l$ and $b_r$. The two quasi-boson operators satisfy
the independent bosonic commutation relations
\begin{eqnarray*}
\lbrack \varphi ,\varphi ^{\dagger }] &=&1,\text{ }[\phi ,\phi ^{\dagger
}]=1, \\
\lbrack \varphi ,\phi ] &=&[\varphi ^{\dagger },\phi ^{\dagger }]=0
\end{eqnarray*}
for very large $N_l$ and $N_r$. This is an approximation up to the order of $%
O(\frac 1{N_s})(s=l,r)$. It is noticed that the ground state ${\bf %
|g_l,g_r\rangle }$ can become the vacuum $|0\rangle $ for $b_l$ and $b_r$, ($%
{\bf |g_l,g_r\rangle =}$ $|{\bf g}_l\rangle $ $\otimes |{\bf g}_r\rangle
\otimes |0\rangle _p$ ${\bf \equiv }|0\rangle _l$ $\otimes |0\rangle
_r\otimes |0\rangle _p$ $=|0\rangle $), and also for $\varphi $ and $\phi $
equivalently.

To show the second observation that $|D(n)\rangle $ is a dark state with
double excitons, we write the effective Hamiltonian as

\begin{eqnarray}
H &=&\omega _aa^{\dagger }a+\omega _lb_l^{\dagger }b_l+\omega _rb_r^{\dagger
}b_r  \nonumber \\
&&+a(g_l\sqrt{N_l}b_l^{\dagger }+g_r\sqrt{Nr}b_r^{\dagger })+{\rm h.c.}
\end{eqnarray}
in the macroscopic limit by ignoring the constant form. When the Stark
shifts are neglected, we have $\omega _l=\omega _r=\epsilon $ and

\begin{equation}
H=\omega _aa^{\dagger }a+\epsilon (\phi ^{\dagger }\phi +\varphi ^{\dagger
}\varphi )+ga\phi ^{\dagger }+{\rm h.c.}.
\end{equation}
In this case, the excitation $\varphi $ decouples with the quantized field
and the complementary excitation $\phi $. It is obvious that double-exciton
dark state $|D(n)\rangle =\frac 1{\sqrt{n!}}\varphi ^{\dagger n}|0\rangle $
can be cancelled by the interaction part $H_I=ga\phi ^{\dagger }+{\rm h.c.}$.

In this limit, we can exactly solve the quantum dynamics for entangling the
collective excitations of atomic ensembles with the quantized light filed.
To this end, we regard this entanglement as a reading-out process of quantum
information stored in the atomic ensemble. To describe the coherent transfer
of the quantum information from atomic ensembles to the quantum light field
in our scheme, we consider the dressed excitation by the bosonic polariton
operators
\begin{eqnarray}
A &=&a\cos \frac \vartheta 2+\phi \sin \frac \vartheta 2,  \nonumber \\
B &=&-a\sin \frac \vartheta 2+\phi \cos \frac \vartheta 2  \label{A,B}
\end{eqnarray}
and rewrite the total Hamiltonian in terms of the canonical modes of
polariton as

\begin{equation}
H=\Theta \widehat{N}+\Xi (A^{\dagger }A-B^{\dagger }B)+\epsilon \varphi
^{\dagger }\varphi .  \label{H(A,B)}
\end{equation}
Here ,
\begin{eqnarray}
\widehat{N} &=&A^{\dagger }A+B^{\dagger }B,  \nonumber \\
\Xi &=&\sqrt{(\frac{\omega -\epsilon }2)^2+g^2}, \\
\Theta &=&\frac{\omega +\epsilon }2,\text{ }\tan \vartheta =\frac{2g}{\omega
-\epsilon }.  \nonumber
\end{eqnarray}

We consider the case that the coupling parameters of the total system do not
change. Let there be no photons in the quantized mode $|0\rangle _p$ and all
the atoms be coherently prepared in the left well with the initial state
\begin{equation}
|S(\eta )\rangle _l=e^{i\eta S_y(l)/\sqrt{N_l}}{\bf |g_l,g}_r{\bf \rangle }%
\otimes |0\rangle _p
\end{equation}
by a collective Rabi rotation. Here, we have defined $S_y(l)=%
\sum_{j=1}^{N_l}\sigma _y^{[j]}(s)$. The atom number
$N_l=\sqrt{\eta }$ possesses a minimum uncertainty. It can be
approximated by a coherent state
\begin{equation}
|\eta \rangle _l=\exp [\eta (b_l^{\dagger }-b_l)]|0\rangle
\end{equation}
in the macroscopic limit, where $|0\rangle =$ $\ |0\rangle _l$ $\otimes
|0\rangle _r\otimes |0\rangle _p$.

In order to calculate the evolution of the state $|\eta \rangle
_l$, we first get the Heisenberg equations of motion of operators
from the Hamiltonian Eq.(\ref{H(A,B)})
\begin{eqnarray}
\stackrel{\bullet }{A}(t) &=&-i(\Theta +\Xi )A(t),  \nonumber \\
\stackrel{\bullet }{B}(t) &=&-i(\Theta -\Xi )B(t), \\
\stackrel{\bullet }{\varphi }(t) &=&-i\epsilon \varphi (t),  \nonumber
\end{eqnarray}
which give
\begin{eqnarray}
A(t) &=&A(0)\exp [-i(\Theta +\Xi )t],  \nonumber \\
B(t) &=&B(0)\exp [-i(\Theta -\Xi )t],  \nonumber \\
\varphi (t) &=&\varphi (0)\exp [-i\epsilon t].
\end{eqnarray}
Together with Eqs.(\ref{phi,psi}) and (\ref{A,B}), one can write down the
operator
\begin{eqnarray}
b_l(t) &=&\varphi (0)\exp [-i\epsilon t]\cos \alpha  \nonumber \\
&&+A(0)\exp [-i(\Theta +\Xi )t]\sin \frac \vartheta 2\sin \alpha \\
&&+B(0)\exp [-i(\Theta -\Xi )t]\cos \frac \vartheta 2\sin \alpha .  \nonumber
\end{eqnarray}
Since the Hamiltonian of the system is bilinear, we have $U(t)|0\rangle
=\exp (-iHt)|0\rangle =|0\rangle $. Therefore, driven by the classical
lights, the total system with the initial state $|\eta \rangle _l$ will
evolve into a factorized state with three approximate components of coherent
state
\begin{eqnarray}
|\eta ,t\rangle &=&U(t)|\eta \rangle _l  \nonumber \\
&=&\exp \{\eta [b_l^{\dagger }(-t)-b_l(-t)]\}|0\rangle  \nonumber \\
&=&\left| \eta f(t)\right\rangle _l\otimes \left| \eta g(t)\right\rangle
_r\otimes \left| \eta h(t)\right\rangle _p,
\end{eqnarray}
where
\begin{eqnarray}
f(t) &=&\cos ^2\alpha \exp (-i\epsilon t)+\sin ^2\frac \vartheta 2\sin
^2\alpha \exp [-i(\Theta +\Xi )t]  \nonumber \\
&&+\cos ^2\frac \vartheta 2\sin ^2\alpha \exp [-i(\Theta -\Xi )t],  \nonumber
\\
g(t) &=&\{-\exp (-i\epsilon t)+\sin ^{2}\frac{\vartheta }{2}\exp
[-i(\Theta
+\Xi )t] \\
&&+\cos ^{2}\frac{\vartheta }{2}\exp [-i(\Theta -\Xi
)t]\}\frac{\sin 2\alpha
}{2},  \nonumber \\
h(t) &=&-i\sin \vartheta \sin \alpha \exp (-i\Theta t)\sin (\Xi t).
\nonumber
\end{eqnarray}
While the second component corresponds to the coherence tunnelling
of the collective internal excitation from the left ensemble to
the right one, the third term presents the evolution of the
quantized light field entangled with the collective excitation of
atomic ensembles. As a consequence, the effective increasing $\eta
^2|\alpha (\varphi ,t)|^2$ of photons in the quantized light mode
can record the information concerning the transfer of collective
excitations in atomic ensembles.

\section{\protect\bigskip Summery}

In conclusion we have discussed the idea of the double-exciton dark states
for two-atomic-ensemble system dressed by a single mode light field. We also
considered the auxiliary role of a quantized light field. The independent
adiabatic manipulation of two classical light fields interacting with two
atomic ensembles is the key technology for the realistic realization.

Obviously, our protocol based on this conception to transfer
information between quantum memories mainly depends on a many-atom
enhancement mechanism to realize a convenient manipulation for the
effective coupling strengths (the two effective coupling strengths
can be changed independently by the stimulated Ramann adiabatic
passage). This is in fact, not quite surprising, that earlier
cavity QED experiments have relied on the enhanced dipole
interaction of a collection of many atoms \cite{sf1,n3,n4}. In
free space, the phenomena of superfluorescence or super-radiance
\cite{sf1} constitutes another example of collective state
dynamics. Notice that the
electric dipole couplings $g_{l}$ $\sim \frac{1}{\sqrt{V_{l}}}$and $g_{r}$ $%
\sim \frac{1}{\sqrt{V_{r}}}$ where $V_{l}$ and $V_{r}$ are the effective
model volume of fields interacting with the left and right ensembles. So the
effective coupling strengths $g_{l}\sqrt{N_{l}}$ and $g_{r}\sqrt{N_{r}}$ are
proportional to the square-root of effective densities $n_{s}=\frac{N_{s}}{%
V_{s}}$ ($s=l,r$) of field-atom interaction. In usual, similar to the
thermodynamics limit, one can think $n_{s}$ as constants since both $N_{s}$
and $V_{s}$ approach infinity. In this sense, the role of many-atom
enhancement in manipulation of the effective coupling strengths is limited
by the density of atomic gas. As mentioned in section 3, the imperfections
of our scheme are due to the inhomogeneous interaction and out-control of
atoms, which may cause the decoherence of order $\sqrt{N_{s}}$. This shows
that the technique is not extremely robust.

\bigskip

{\it The present work is supported by the CNSF (grant No.90203018)
and the knowledged Innovation Program (KIP) of the Chinese Academy
of Sciences and the National Fundamental Research Program of China
with No 001GB309310.  We also sincerely thank L. You for the
helpful discussions with him.}

\end{document}